\documentclass[12pt]{article}
\usepackage{amstex,epsfig}
\def\g5{\gamma^5}

\def\d4k{{d^4k\over (2\pi)^4}}

\def\q{\rm{q}}
\def\barq{\rm{\bar q}}
\def\u{\rm{u}}
\def\baru{\rm{\bar u}}
\def\d{\rm{d}}
\def\bard{\rm{\bar d}}
\def\qq{<\rm{\bar q q}>}
\def\GG{<{\rm G^2}>}
\newcommand{\beq}{\begin{eqnarray}}
\newcommand{\eeq}{\end{eqnarray}}
\begin{document}
\title{Scalar Mesons, Glueballs, Instantons and the Glueball/Sigma}

\author{Leonard S. Kisslinger\\
        Department of Physics,\\
       Carnegie Mellon University, Pittsburgh, PA 15213\\
                    and\\
      Mikkel B. Johnson\\
      Los Alamos National Laboratory, Los Alamos, NM 47545}

\maketitle
\indent
\begin{abstract}
We include instanton effects in QCD sum rules for coupled scalar glueballs
and mesons. We find a light glueball/sigma as in earlier studies without 
instantons, but in a lattice-type pure instanton model the light
glueball/sigma is not found.  In the 1-2 Gev region we now find
that lightest I=0 meson, in the region of the f$_o$(1370), has no direct
glueball mixing, with the instanton loop replacing the glueball component.
The lightest scalar mainly glueball in the region of the f$_o$(1500) 
is sensitive to the choice of nonperturbative gluonic parameters.
\end{abstract}

\vspace{0.5 in}

\noindent
PACS Indices:12.38.Lg,14.40.-n,14.70.Dj,13.75.Lb

\newpage
\section{Introduction}
\hspace{.5cm}

  Some of the earliest applications of the method of QCD sum rules was for
the study of gluonic hadrons\cite{nsvz}, known as glueballs. Using the sum
rules obtained in that work it was observed that a light scalar glueball
solution in the region of 400-500 MeV~\cite{lk1} might be strongly coupled to
a $\pi\pi$ resonance we call the sigma, which would give large branching 
ratios for the decays into channels with sigmas of heavier glueballs and 
hybrids\cite{lk2}. Moreover, it was observed\cite{lk1} that if the coupling
of scalar mesons to glueballs is included, the original scalar meson QCD  
sum rules\cite{ryr} are modified so that the lightest 80\% meson solution 
predicts a mass about 400 MeV higher than the pure $\q\barq$ solution,
{\i.e.}, near the f$_o$(1370) rather than the f$_o$(980) as found in the
earlier work\cite{ryr}

  It has been known for decades that instantons can represent a large
part of nonperturbative gluonic interactions. In the present work we
include instanton effects in the sum rules for the scalar glueballs
and mixed scalar mesons and glueballs.  In Sect.~2.1 we review
the QCD sum rules for scalar mesons, glueballs and mixed meson-glueballs
without instantons. In Sect.~2.2, we review work that has been
done on scalar hadrons with instantons using the sum rule methods and
give the modification of the sum rules with instantons included. In 
Sect.~3 we give the results for a possible light glueball/sigma of mass
about 500 MeV and for mixed meson-glueballs in the 1-2 GeV region.  If we drop 
all gluonic condensates from the sum rules, but retain the instantons as
the source of nonperturbative effects to agree with recent quenched lattice 
calculations, 
we no longer find the light glueball.  Instead, we find only scalar glueball 
solutions with masses greater than 1400 MeV.  We discuss
the results in Sect.~4.

\section{Sum Rules for Mixed Scalar Mesons and Glueballs}

\vspace{.5 cm}

In this section we discuss the QCD sum rules for mixed scalar mesons and
quarks. The method\cite{svz} makes use
of a correlator defined in terms of a composite field operator
\beq
\label{1}
\Pi (p) & = & i\int d^4 x \; e^{iq \cdot x} <0 \mid T [J(x) J(0)] \mid 0> \; ,
\eeq
where J(x) is a field operator composed of quark and/or gluon fields for
pure QCD. In the present work the operators are
\beq
\label{2}
            J^m (x) &  = & \frac{1}{2}(\baru (x) \u (x)- \bard (x) \d (x)) \\
 \nonumber
            J^G (x) & = & \alpha_s G^2,
\eeq
for the I=0 scalar meson and glueball, respectively. The I=1 scalar meson
is not treated here. The QCD sum rules are
obtained by equating a dispersion relation for the correlator to a
QCD evaluation using an operator product expansion (OPE). We discuss 
these sum rules in the following two subsections, first without explicit 
instantons and secondly with explicit instanton contributions.

\subsection{Mixed Scalar Mesons and Glueballs Without Instantons}

In this subsection we review QCD sum rules for scalar mesons, for
scalar glueballs, and for the mixed meson-glueball sum rules. As was discussed
in the first QCD sum rule research on scalar glueballs\cite{nsvz}, 
there is a strong argument for using a subtracted dispersion relations,
while the first work on scalar mesons\cite{ryr} used an unsubtracted
dispersion relation. This is important consideration for the present
work.

\subsubsection{Scalar mesons}

The QCD sum rules for scalar mesons were first treated in Ref.~\cite{ryr}.
The most important processes, obtained by an OPE, are illustrated in Fig. 1
\begin{figure}
\begin{center}
\epsfig{file=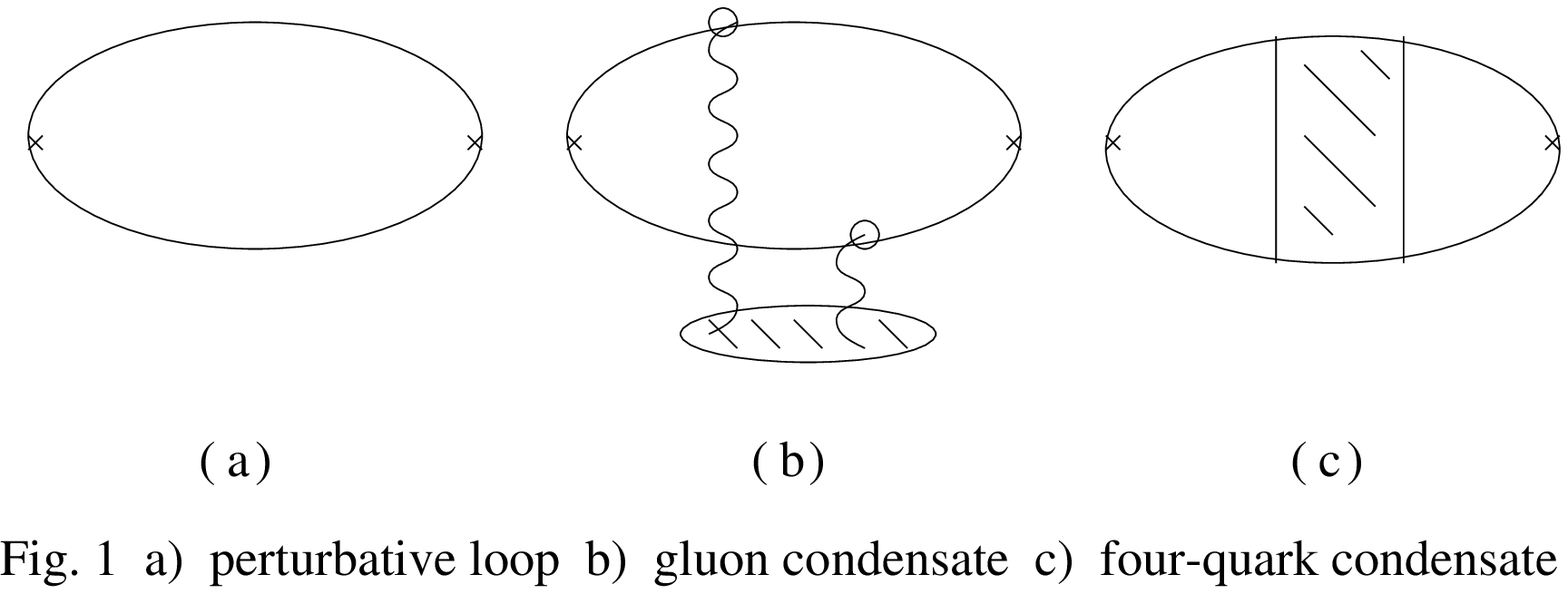,height=5cm,width=12cm}
{\label{Fig.1}}
\end{center}
\end{figure}
for light-quark mesons, and in Euclidean momentum space (Q$^2$=-p$^2$) give
\beq
\label{3}
  \Pi (Q)^{QCD} & = & \frac{3}{8 \pi^2} (1+\frac{11}{3} \frac{\alpha_s}{\pi})
 Q^2 ln (Q^2) \\ \nonumber
                  &   & +\frac{\alpha_s}{8 \pi Q^2} \GG
 +\frac{ \pi \alpha_s}{Q^4} P^{4q},
\eeq
where $ \GG $ is the gluon condensate and $P^{4q}$ is the
four-quark condensate for the scalar meson, illustrated in Fig. 1c. 
The perturbative gluon correction diagram is not shown. Note 
that the quark condensate term, $\qq$, is neglected as it is
proportional to the current quark mass. The factorized form For the 
four-quark condensate~\cite{svz}, $P^{4q} \simeq -176 \qq^2/27$, is
probably accurate to about a factor of two. After the Borel transform the 
sum rule, with M the Borel mass, is
\beq
\label{4}
 g_0 e^{-M_m^2/M^2} & = & \frac{3}{8 \pi^2}(1+\frac{11 \alpha_s}{3 \pi})
 M^4 E_1(s_o/M^2) \\ \nonumber
                    &   &+ \frac{\alpha_s}{8 \pi} \GG 
-\frac{176 \alpha_s}{27 M^2} c_4 \qq^2,
\eeq
with c$_4$ a constant representing the correction to the four-quark
condensate, and g$_0$ a D = 4 constant which will not be used in the
analysis. The functions E$_n$(s$_0/M^2$) represent the continuum contribution
to the sum rules with a simple form for the continuum spectral function.
They are defined as $ E_n(x) = 1 - e^{-x} \sum_{k=0}^{n} \frac{x^k}{k!})$

  Taking the derivative of Eq.(\ref{4}) with respect to 1/M$^2$ and using
the ratio of the resulting equation and  Eq.(\ref{4} one obtains an
equation for the scalar meson mass:
\beq
\label{5}
 M_m^2 & = & \frac{\frac{3}{8 \pi^2}(1+\frac{11 \alpha_s}{3 \pi})
 (2M^6 E_1(s_0/M^2)-s_o^2M^2e^{-\frac{s_o}{M^2}})-\pi \alpha_s P^{4q}}
     {\frac{3}{8 \pi^2}(1+\frac{11 \alpha_s}{3 \pi})
 M^4 E_1(s_0/M^2)+\frac{\alpha_s}{8 \pi} \GG +\frac{\pi \alpha_s}{M^2}P^{4q}}
\eeq

There is a stable solution for the scalar meson mass M$_m$ of about 1 GeV, which
is the result of the original work of Ref.\cite{ryr}. If this were the
physical solution then we would interpret this as the f$_0$(980), however,
it was found in previous work~\cite{lk1} that the coupling to the scalar 
glueball increases the mass by 3-400 MeV. 
In the simple picture of Eq.(\ref{5}) the
I=1 is degenerate with the I=0 meson, giving the a$_0$(980).

\subsubsection{Scalar glueballs}

The QCD sum rules for scalar glueballs were first treated in 
Ref.~\cite{nsvz}. There have been many calculations of scalar
glueballs in recent years~\cite{gb,lk1}. With the preturbative corrections 
of Ref.\cite{bs} and including nonperturbative terms up to dimension eight
one finds for the correlator corresponding to the processes 
illustrated in Fig. 2
\begin{figure}
\begin{center}
\epsfig{file=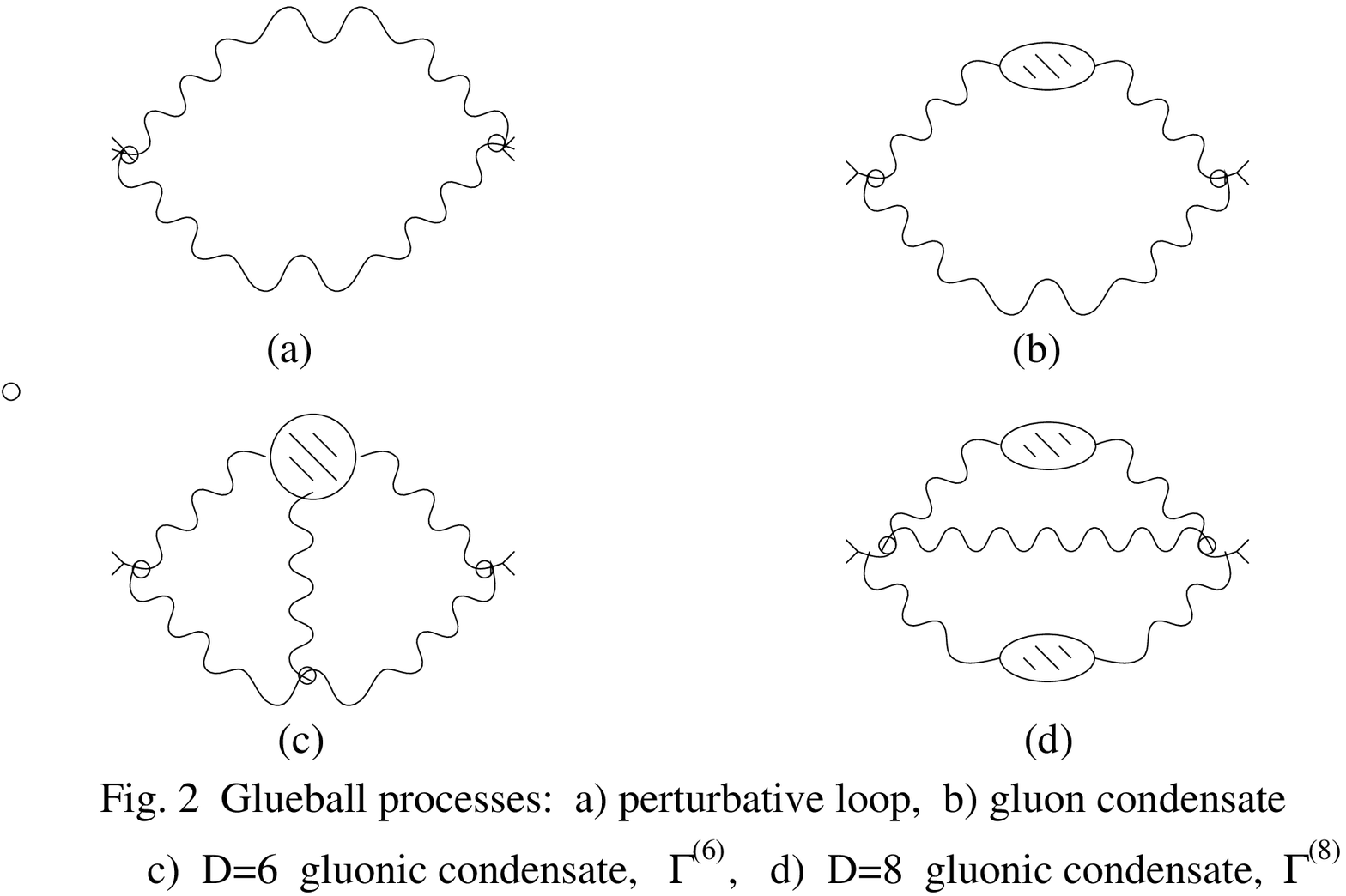,width=12cm}
{\label{Fig.2}}
\end{center}
\end{figure}
\beq
\label{6}
    \Pi (Q^2)^{QCD} & = &  -2 (\frac{\alpha_s}{\pi})^2(1+ \frac{51}{4}
 \frac{\alpha_s}{\pi} -\frac{11}{4} \frac{\alpha_s}{\pi} ln(Q^2))
 Q^4 ln(Q^2) +4 \alpha_s^2 \GG (1+ \\ \nonumber
       & &  \frac{49}{12} \frac{\alpha_s}{\pi} 
  -\frac{11}{4} \frac{\alpha_s}{\pi} ln(Q^2))
      + \frac{8 \alpha_s^2 \Gamma^{(6)}}{Q^2} (1- \frac{29}{4} \alpha_s
 ln(Q^2)) +  \frac{8 \pi \alpha_s^3 \Gamma^{(8)}}{Q^4},
\eeq
where $\Gamma^{(6)}$ =
 $\langle g_s f_{abc} G^a_{\mu\nu} G^b_{\nu\rho} G^c_{\rho\mu} \rangle$ and
$\Gamma^{(8)}$ =  $\langle 14(f_{abc} G^a_{\mu\nu} G^b_{\nu\rho})^2
-(f_{abc} G^a_{\mu\nu} G^b_{\rho\lambda})^2  \rangle$ are dimension 6 and
dimension 8 gluonic condensates, illustrated in Figs 2c) and d), respectively.
Note that the largest nonperturbative contribution from the
gluon condensate is independent of momentum, since the dimension
of $\GG$ is the same as the correlator. For this reason it does not contribute
to the correlator after the Borel transform. For this reason it was 
suggested~\cite{nsvz} that one should use a subtracted dispersion relation.
Using the subtracted form, $(\Pi(Q^2) - \Pi(0))/Q^2$ and taking the Borel
transform one obtains the sum rule
\beq
\label{7}
  \Pi(0) e^{-M_G^2/M^2} + cont. & = &\Pi(0) 
  +2 (\frac{\alpha_s}{\pi})^2(1+ \frac{51}{4} \frac{\alpha_s}{\pi} 
 -\frac{11}{2} \frac{\alpha_s}{\pi} (1-\gamma_E+ln(M^2))M^4
  \\ \nonumber
     & &  E_1(s_o/M^2)- 4 \alpha_s^2 \GG(1+ \frac{49}{12} 
 \frac{\alpha_s}{\pi} +\frac{11}{4} \frac{\alpha_s}{\pi}(\gamma_E -ln(M^2)) 
 \\ \nonumber
    &  & - \frac{8 \alpha_s^2 \Gamma^{(6)}}{M^2} (1- \frac{29}{4} \alpha_s
 (1-\gamma_E +ln(M^2))) - \frac{8 \pi \alpha_s^3 \Gamma^{(8)}}{2 M^4}.
\eeq
Taking the ratio of the $\partial/\partial(1/M^2$) of Eq.(\ref{7}) to
Eq.(\ref{7}) one has a sum rule for a pure scalar glueball, with no
explicit quark/antiquark components, although the qluonic condensates
contain important quark pair contributions,
\beq
\label{8}
 M_G^2 & = & 
  [2 (\frac{\alpha_s}{\pi})^2 (2 (1+ \frac{51}{4} \frac{\alpha_s}{\pi}
 -\frac{11}{2} \frac{\alpha_s}{\pi} (1-\gamma_E +ln(M^2))) M^6 E_2(s_o/M^2)
     \\ \nonumber
   & &-\frac{11}{2} (\frac{\alpha_s}{\pi}) M^6 E_1(s_o/M^2))
  + 11 (\frac{ \alpha_s^3}{\pi}) \GG M^2
  + 8 \alpha_s^2 \Gamma^{(6)} \\ \nonumber
  & & (1+ \frac{29}{4} \alpha_s (\gamma_E -ln(M^2)) 
+ \frac{8 \pi \alpha_s^3 }{M^2} \Gamma^{(8)}] \\ \nonumber
  & & [\Pi(0) +2 (\frac{\alpha_s}{\pi})^2  
   (1+ \frac{51}{4} \frac{\alpha_s}{\pi}
 -\frac{11}{2} \frac{\alpha_s}{\pi} (1-\gamma_E +ln(M^2)))M^4 E_1(s_o/M^2) \\
\nonumber & & - 4 \alpha_s^2 
\GG (1+ \frac{49}{12}\frac{\alpha_s}{\pi} +\frac{11}{4} \frac{\alpha_s}{\pi}
  (\gamma_E  -ln(M^2)) \\ \nonumber
& &  - \frac{8 \alpha_s^2}{M^2} \Gamma^{(6)} (1
- \frac{29}{4} \alpha_s
 (1-\gamma_E +ln(M^2))) - \frac{4 \pi \alpha_s^3}{M^4}
\Gamma^{(8)}]^{-1}.
\eeq
Using the range of $\Gamma^{(6)}$ and $\Gamma^{(8)}$ suggested in 
Refs.\cite{gb,bs} one finds stable solutions for a light glueball with a 
mass in therange of 300 MeV-600 MeV.  This is the glueball/sigma of 
Ref.~\cite{lk1}, which has been used for calculating the sigma 
branching ratio from hybrid decay\cite{lk2} and estimates of diffractive
sigma production in high energy proton-proton processes mediated by the 
Pomeron\cite{kms}.

\subsubsection{Mixed scalar glueballs and mesons}

  Physically it is expected that scalar glueballs and mesons should mix,
and from the early calculations using QCD sum rules for scalar 
hadrons\cite{nsvz,nar} this was discussed.
This suggests that one must use a scalar current of the form
\beq\label{9}
 J_{0^{++}} & = & \beta M_o J_m + (1-\mid\beta\mid) J_G,
\eeq
where M$_o$ is a constant that we take to be 1 GeV. The
mechanism for mixing is given by a low energy theorem discussed 
in Ref.~\cite{nsvz}.  One can show that the
mixing term in the correlator with the current of Eq.(\ref{9}) is~\cite{lk1}
\beq
\label{10}  
    \Pi^{mixing} & \simeq & \beta (1-\mid\beta\mid)\frac{64}{9} \qq.
\eeq
Since the Borel transforn of a constant vanishes, the mixing term does
not contribute to the unsubtracted sum rule, while it contributes the
term shown in Eq.(\ref{10}) to the subtracted dispersion relation.
One finds the solutions to the sum rules discussed in Ref.~\cite{lk1},
where it was reported that there 
are no stable solutions in the 1-2 GeV region without glueball-meson
mixing. The most stable solutions were found for a 80\% meson and an 80\%
glueball. These presumably correspond to the f$_o$(1370) and f$_o$(1500),
respectively, although the uncertainty of about 15\% in the solutions
cannot separate these two solutions. Note that there is no solution
near the f$_o$(980), which shows the importance of the gluonic mixing
(recall that without the mixing the lowest mesonic solution is near the
f$_o$(980)\cite{ryr}). The purely gluonic solution in the 300-600 MeV
range is a prediction of the method.  In the next sections we study the
solutions to the sum rules with instanton effects included.

\hspace{.5cm}

\subsection{Mixed Scalar Mesons and Glueballs With Instantons}

\vspace{.5 cm}

   For many years it has been known that instantons can represent a
major part of the nonperturbative gluoic interactions. See Ref.~\cite{ss}
for an excellent review of the concepts of instantons and applications 
to QCD. The starting point is the solution for the instanton using
the classical action~\cite{bel}, which gives for the instanton color
field
\beq
\label{11}
   A^{inst}_{\mu(x) a} & = & \frac{2 \eta_{a\mu\nu} x_\nu}{x^2 + \rho^2} 
 \\ \nonumber
     G^{inst}(x) \cdot G^{inst}(x) & = & \frac{192 \rho^4}{(x^2 + \rho^2)^4}
\eeq
where $\rho$ is the instanton size.  From this the quark zero modes were
derived~\cite{th}, which is the main basis for subsequent research with
quarks and instantons.  Although in the instanton gas 
picture, with $\rho \simeq 1.0$ fm\cite{nsvz}, the inclusion of 
instantons in QCD sum rules did not seem very promising for hadronic 
physics, in the instanton liquid model~\cite{shu1} with
$\rho \simeq 1/3$ fm instantons give large nonperturbative effects
in the medium-range region, where neither the perturbative glue nor 
the long range confining glue represented by the condensates are
effective. On the other hand the other hand the instantons cannot
give confinement (see Ref.\cite{ss} for a discussion and references).
From this we conclude that one needs both instanton and gluonic
condensate processes in the QCD sum rules for scalar hadrons.

In a previous study of the role of instantons in hadronic physics
the solution for a quark in the instanton-antiinstanton
medium~\cite{pol} was tested using the Dyson-Schwinger (DS) 
equation~\cite{lk3} with a
confining gluonic propagator that was fit to the condensates, and 
it was found that there is no consistency with the condensates. 
An important observation for the present work is that although the
instantons can produce most of the quark condensate (dimension = 3)
they do not give the correct higher dimensional mixed condensate
(dimension = 5). This can be explained by the higher dimensional
condensates produced by gluonic effects at a larger length scale
than the 1/3 Fm of the instanton liquid model.
Recently, the DS equation was solved both on the light-cone~\cite{kl} 
and in four-dimensional Euclidean space~\cite{hk}, and consistent 
solutions are obtained 
if both instantons and a confining gluonic propagator are included.
Moreover, the instantons provide the largest nonperturbative effects.

In the present work we assume that there are three length scales: the
perturbative region with L$\leq$ 0.2 fm, the midrange nonperturbative
region with L$\simeq$ 0.33 fm given by instantons and the confining
region with L$ \geq$ 0.5 fm. Our model is as follows:\\

  1) The color field can be written as A = A$^{inst}$ + $\bar{\rm{A}}$,
with the instanton being the classical solution and  $\bar{\rm{A}}$ the
residual color quantum field. 

  2) The instanton loop is included in the correlator for scalar glueballs,
but no instanton interactions are included. The latter are assumed to
be accounted for by the various condensates. This is consistent with the
instanton liquid model~\cite{ss}, in which the paramenters are constrained
by the gluon condensate.

  3) In the meson correlator the loop of quark-antiquarks in the background
instanton medium is included, but no instanton interacton processes, for
similar reasons.\\

Thereby, we add the instanton processes to
the perturbative processes for the sum rules.
This adds two processes to the QCD side of the correlator.
For the scalar meson correlator the additional process
is the loop with a quark and an antiquark in the instanton medium.
After a Borel transform the instanton contribution
to an isospin = I scalar meson correlator is~\cite{shu2,deks,hf}
\beq
\label{12}
  \Pi^{\q\barq,inst} & = & (-1)^I \frac{3}{8 \pi^2} \rho^2 M^6 e^{-x} 
 (K_o(x) +K_1(x)),
\eeq
with x=$\rho^2 M^2/2$. The instanton continuum contribution corresponding this
is
\beq
\label{13}
  \Pi^{\q\barq,inst, cont} & = & (-1)^I \frac{3}{4 \pi} \int_{s_o}^{\infty}
 ds s J_1(\rho \sqrt{s}) Y_1(\rho \sqrt{s}) e^{-s/M^2},
\eeq
with s$_o$ the continuum parameter. The notation for the various Bessel
functions is standard~\cite{as}.

  The new contribution of the instanton loop for the scalar glueball
for the unsubtracted correlator, such
as that used in Eq.(\ref{7}), is (after the Borel 
transform)~\cite{nsvz,shu1}
\beq
\label{14}
   \Pi^{GB,inst}(M) & = & -2^7 \pi^2 n \frac{x^2}{\rho^2} e^{-x} 
  (2 x^3 K_o(x) + (x^2 + 2 x^3) K_1(x)),
\eeq
where $n$ is the instanton density.  The corresponding continuum contribution
is
\beq
\label{15}
 \Pi^{GB,inst,cont}(M) & = & 2^4 \pi^3 n \rho^4 \int_{s_o}^{\infty} ds s^2 
 J_1(\rho \sqrt{s}) Y_1(\rho \sqrt{s}) e^{-s/M^2}.
\eeq
For the subtracted dispersion relationship, which gave the solution for the
light glueball/sigma discussed above, the QCD and continuum contributions
are
\beq
\label{16}
   \Pi^{GB,inst}(M) & = & 2^6 \pi^2 n x^2 e^{-x} 
  ((1+x) K_o(x) + (2 + x + \frac{2}{x}) K_1(x))-2^7 \pi^2 n
\eeq
for the instanton loop and 
\beq
\label{17}
 \Pi^{GB,inst,cont}(M) & = & 2^4 \pi^3 n \rho^4 \int_{s_o}^{\infty} ds s
 J_1(\rho \sqrt{s})
 Y_1(\rho \sqrt{s}) e^{-s/M^2}.
\eeq
from the continuum.

\section{Results}

\subsection{Light Scalar Glueball, Lattice Gauge Comparison}

We now search for a solution to the subtracted correlator sum rules, which
in the previous work without instantons~\cite{lk1} yielded the light scalar 
glueball/sigma. The sum rule
solution is given by Eq.(\ref{8}) with the addition of the instanton terms
given by Eqs.(\ref{16},\ref{17}), where the suitable derivative of the
instanton terms must be carried out for the numerator of the equation.
A crucial question is that with the inclusion of instantons how should one
modify the values of the dimennsion 4, 6, and 8 gluonic condensates,
$<G^2>, \Gamma^{(6)}$, and $\Gamma^{(8)}$, discussed above. 
An analysis of recent
QCD lattice calculations~\cite{jn} finds that these quenched calculations give 
results similar to the instanton liquid model. However, this instanton
model does not give the correct string tension. The interpretation must be
that the instantons can give the quark condensate, where the length scale is
about 1/3 fm, but cannot give the infrared nonperturbative QCD (NPQCD) 
effects. We explore this in the following manner: first we look for solutions 
with instantons and the gluonic condensates, studying the solutions with 
suitable choices of the
gluonic condensates. We then look for solutions with the only NPQCD
effects being those given by instantons, which should resemble the 
lattice gauge calculations~\cite{jn}. 

We do indeed find  solutions to the subtracted dispersion relation using 
higher-dimension gluonic condensates in the range previously found and with the 
standard parameters of the instanton liquid model~\cite{ss}. Specifically, the 
size
of the instanton is 1.67 GeV$^{-1}$ and the instanton density is .0008 GeV$^4$.
A typical solution as a function of the Borel mass is shown in Fig.~3.
\begin{figure}
\begin{center}
\epsfig{file=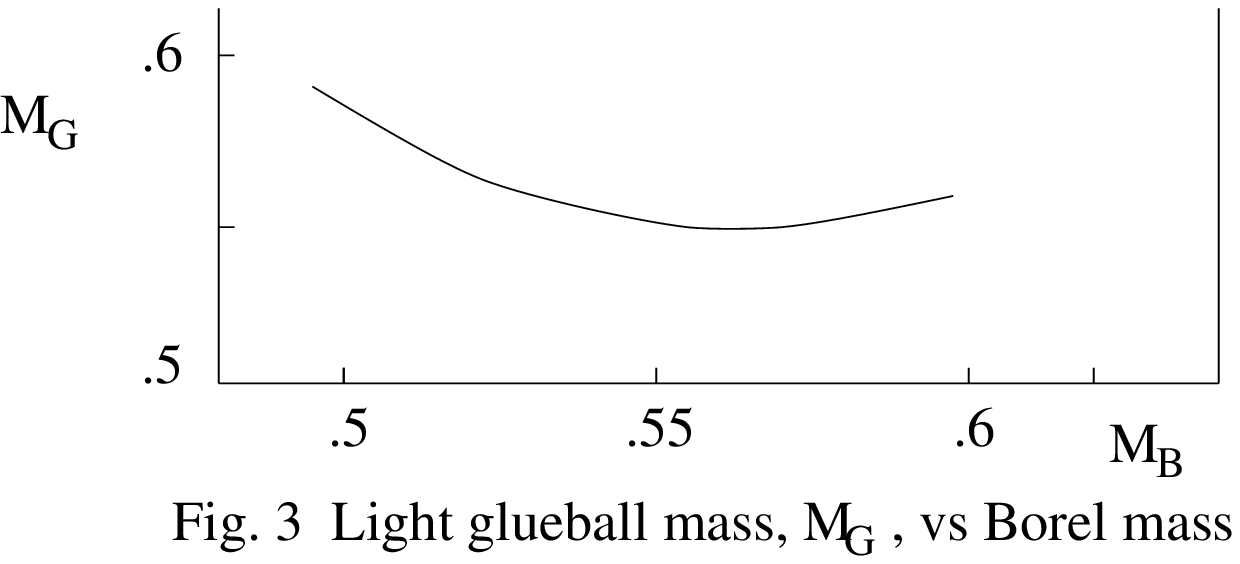,width=12cm}
{\label{Fig.3}}
\end{center}
\end{figure}
The greatest uncertainty in the method is the values of the gluonic
condensates, particularly the higher-dimensional condensates,
$\Gamma^{(6)}$ and $\Gamma^{(8)}$. 
The most widely accepted values of these higher dimensional gluonic
condensates are $\Gamma^{(6)}$ = .0114 GeV$^6$ and $\Gamma^{(8)}$
 = .0081 GeV$^8$.
The solution shown in Fig.~3 has values of these condensates reduced by
20\%.
We also find stable solutions that meet the criteria of QCD sum rules
in the range 400-600 MeV for values of the higher-dimensional condensates
40\% to 100\% of the accepted values mentioned.
Fortunately,
the ordinary gluonic condensate does not play a major role in these
calculations once the instantons are included.

Next we carry out studies of the subtracted dispersion dropping all the
gluonic condensates, which should give solutions similar to the
quenched lattice calculations~\cite{jn}, if our arguments are correct. 
No light
glueball solution is found, as in the present lattice calculations.
The solutions are indeed quite sensitive to the instanton parameters.
A stable solution for a glueball with a mass of about 1560 MeV is shown
in Fig.~4, with parameters instanton size =  1.28 GeV$^{-1}$
\begin{figure}
\begin{center}
\epsfig{file=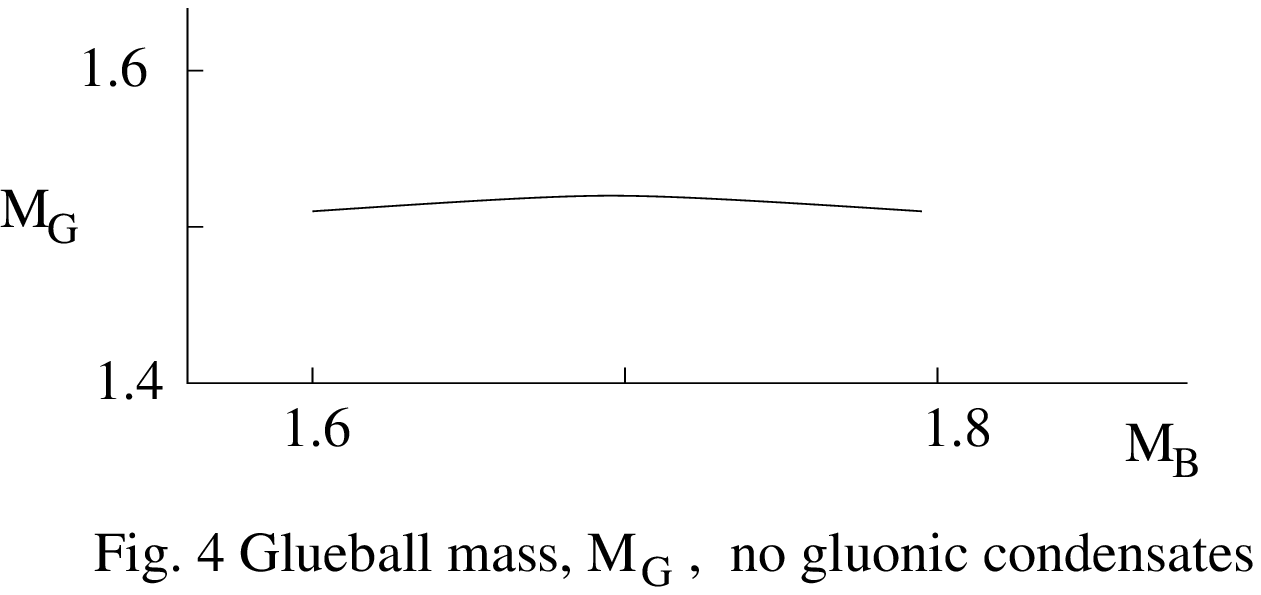,width=12cm}
{\label{Fig.4}}
\end{center}
\end{figure}
and density = .00018 GeV$^4$. If one uses instanton parameters closer
to or equal to the standard ones given above the glueball solutions
are heavier, but we do not find consistency in the sense of the
mass range of the Borel plateau of the sum rules including the mass
of the solution. In other words with a pure gluonic instanton model
we do not find stable solutions for a scalar glueball. This is 
consistent with our observations that in the 1-2 GeV regions the
stable glueballs must have a scalar meson component.
Note that recent lattice glueball solutions find
the lightest scalar glueball mass at about 1700 MeV~\cite{cm,bali}.

We conclude that stable scalar glueball solutions in the range of 
400-600 MeV are found if one includes the higher-dimensional gluonic 
condensates, and to the extent that the sum rule method is reliable 
there is a light glueball that could be part of the strongly
coupled glueball/sigma system.  Moreover, we find a plausable explanation 
of the present lattice calculations not finding a light glueball.

\hspace{.5cm}

\subsection{Mixed Scalar Mesons}

  For the solutions to the unsubtracted sum rules our most striking new result 
in the 1-2 GeV region is that there are now, with instantons, no stable 
solutions in which there is a dominant scalar meson ($\q\barq$)
component with an even small admixture of a scalar glueball. 
This is in contrast
to our earlier result~\cite{lk1} in which the only solutions obtained after
the inclusion of the mixing given in Eq.(\ref{10})
had a mixing of the order of 20 \%. The process of
the loop of $\q\barq$ in which the propagators are those of quarks in the
background instanton-antiinstanton medium seems to replace the physical
input of the glueball. This seems to be physically reasonable, but is not
obvious. We obtain a second very stable solution with 80\% glueball and
20\% meson, which is in the region of the f$_o$(1500); however, in contrast
to the purely meson solution, the mainly glueball solution depends strongly
on the choice of parameters.  In fact, for the standard instanton liquid
model there are no consistent solutions.

\section{Discussion}

 For the subtracted sum rules, which are given by Eq.(\ref{8}) with the
addition of the instanton contributions to the glueball correlator, given by 
Eqs.(\ref{16},\ref{17}), we find  good solutions in the 400-600 MeV region
as has been found in Refs.\cite{gb,lk1,bs} without instantons. The instanton
terms play a major role, and our solutions are more stable than those
found earlier\cite{lk1} and as expected less sensitive to the values of
the higher-order gluonic condensates $\Gamma^{(6)}$ and $\Gamma^{(8)}$.
We also have shown that if one only includes instantons without 
higher-dimensional gluon condensates there are no
solutions for light glueballs. Moreover, we find no stable self-consistent
scalar glueballs with the parameters of the instanton liquid model without
scalar meson admixing. This suggests that when accurate unquenched lattice
calculations are carried out that successfully account for the higher-dimension
gluonic condensates they should also find a scalar glueball in the region
of the sigma, about 500 MeV. Our results for a glueball in the 500 MeV region
anddifficulty in obtaining consistent solutions in the 1-2 GeV region are
consistent with the work of Ref.~\cite{bs}.  Recently, a calculation~\cite{hf}
using a double subtracted dispersion relation with both instantons and gluonic
condensates find a scalar glueball solution in the 1.5 GeV region.

      If our sigma/glueball conjecture is correct it
makes the experimental study of branching ratios with channels containing
sigmas for various glue-rich processes, such as hybrid decays and reactions
dominated by Pomeron exchange, as well as heavy meson decays\cite{lk4} most 
interesting.   

\vspace{.5 cm}

  This work was supported in part by the NSF grant PHY-00070888 and in part 
by the DOE contract W-7405-ENG-36. LSK wishes to thank the TQHN group at the
University of Maryland for hospitality during the period when this work was
completed and Colin Morningstar for helpful discussions.

\end{document}